\pdfoutput=1

\documentclass[11pt,twoside,a4paper,cmspaper,final,collab]{cms-tdr}
\def\svnVersion{255588}\def\cmsCernNo{2014-022}\def\cmsCernDate{\today}\def\cmsMessage{Published in the European Physical Journal C as \href{http://dx.doi.org/10.1140/epjc/s10052-014-2973-5}{\doi{10.1140/epjc/s10052-014-2973-5.}}}

\begin{document}\cmsNoteHeader{SMP-13-011}

\hyphenation{had-ron-i-za-tion}
\hyphenation{cal-or-i-me-ter}
\hyphenation{de-vices}

\RCS$Revision: 247065 $
\RCS$HeadURL: svn+ssh://svn.cern.ch/reps/tdr2/papers/SMP-13-011/trunk/SMP-13-011.tex $
\RCS$Id: SMP-13-011.tex 247065 2014-06-18 17:37:46Z alverson $
\providecommand{\cV}{\ensuremath{\cmsSymbolFace{V}}\xspace}

\newcommand{\BDT}{BDT\xspace}
\newcommand{\mbb}{\ensuremath{m_{\bbbar}}}
\newcommand{\mH}{\ensuremath{m_{\PH}}}
\newcommand{\WToLN}{\ensuremath{\PW^{\pm}\to\ell^{\pm}\cPgn}}
\newcommand{\WtoEN}{\ensuremath{\PW\to\Pe\cPgn}}
\newcommand{\WtoTN}{\ensuremath{\PW\to\Pgt\cPgn}}
\newcommand{\WtoMN}{\ensuremath{\PW\to\Pgm\cPgn}}
\newcommand{\ZtoBB}{\ensuremath{\cPZ\to\bbbar}}
\newcommand{\ZToNN}{\ensuremath{\cPZ\to\cPgn\cPagn}}
\newcommand{\ZToLL}{\ensuremath{\cPZ\to\ell^{+}\ell^{-}}}
\newcommand{\ZtoLL}{2-lepton}
\newcommand{\WtoLN}{1-lepton}
\newcommand{\ZtoNN}{0-lepton}
\newcommand{\ZtoMM}{\ensuremath{\cPZ\to\MM}\xspace}
\newcommand{\ZtoEE}{\ensuremath{\cPZ\to\EE}\xspace}
\newcommand{\Vbb}  {\ensuremath{\cV\bbbar}}
\newcommand{\ptjj}  {\ensuremath{{\pt}^{\bbbar}}}
\newcommand{\MZ}{\ensuremath{M_{\cPZ}}}
\newcommand{\dRJJ}{\ensuremath{\Delta R(\bbbar)}}
\newcommand{\CSVmax}{\ensuremath{\mathrm{CSV}_{\text{max}}}}
\newcommand{\CSVmin}{\ensuremath{\mathrm{CSV}_{\text{min}}}}
\newcommand{\dphiVZ}{\ensuremath{\Delta\phi(\cV,\cPZ)}}
\newcommand{\dphiMJ}{\ensuremath{\Delta\phi(\MET,\text{jet})}}
\newcommand{\dphiMtrk}{\ensuremath{\Delta\phi(\MET,{\MET}^{\text(trks)})}}
\newcommand{\ptV}{\ensuremath{\pt^{\cV}}}
\newcommand{\Naj}{\ensuremath{N_{\mathrm{aj}}}}
\newcommand{\ppWZ}{\ensuremath{\sigma(\Pp\Pp \to \PW\cPZ)}}
\newcommand{\ppZZ}{\ensuremath{\sigma(\Pp\Pp \to \cPZ\cPZ)}}
\newcommand{\muBDT}{\ensuremath{\mu = 1.09 {}_{-0.21}^{+0.24}}}
\newcommand{\muMbb}{\ensuremath{\mu = 0.97 {}_{-0.29}^{+0.32}}}
\newcommand{\muWZ}{\ensuremath{\mu_{\PW\cPZ} = 1.37 {}_{-0.37}^{+0.42}}}
\newcommand{\muZZ}{\ensuremath{\mu_{\cPZ\cPZ} = 0.85 {}_{-0.31}^{+0.34}}}
\newcommand{\XSZZ}{\ensuremath{\sigma (\Pp\Pp \to \cPZ\cPZ) = 6.5 \pm1.7\stat \pm 1.0\syst \pm 0.9\thy \pm 0.2\lum\unit{pb}}}
\newcommand{\XSWZ}{\ensuremath{\sigma (\Pp\Pp \to \PW\cPZ) = 30.7 \pm 9.3\stat \pm 7.1\syst \pm 4.1\thy \pm 1.0\lum\unit{pb}}}
\newcommand{\XSZZfid}{\ensuremath{\sigma (\Pp\Pp \to \cPZ\cPZ) = 0.90 \pm 0.23\stat \pm 0.16\syst\unit{pb}}}
\newcommand{\XSWZfid}{\ensuremath{\sigma (\Pp\Pp \to \PW\cPZ) = 4.8 \pm 1.4\stat \pm 1.1\syst\unit{pb}}}
\newcommand{\theoryXSWZ}{\ensuremath{\ppWZ = 22.3 \pm 1.1\unit{pb}}}
\newcommand{\theoryXSZZ}{\ensuremath{\ppZZ = 7.7 \pm 0.4\unit{pb}}}
\newcommand{\theoryXSWZfid}{\ensuremath{\ppWZ = 3.39 \pm 0.17\unit{pb}}}
\newcommand{\theoryXSZZfid}{\ensuremath{\ppZZ = 1.03 \pm 0.05\unit{pb}}}

\cmsNoteHeader{SMP-13-011} 
\title{Measurement of WZ and ZZ production in pp collisions at $\sqrt{s} = 8\TeV$ in final states with b-tagged jets}

\date{\today}

\abstract{Measurements are reported of the WZ and ZZ production cross sections in proton-proton collisions at $\sqrt{s} = 8$\TeV in final states where one Z boson decays to b-tagged jets. The other gauge boson, either W or Z, is detected through its leptonic decay (either $\PW\to\Pe\Pgn$, $\Pgm\Pgn$ or $\cPZ\to\Pep\Pem$, $\Pgmp\Pgmm$, or $\cPgn\cPagn$).  The results are based on data corresponding to an integrated luminosity of 18.9\fbinv collected with the CMS detector at the Large Hadron Collider. The measured cross sections, $\sigma (\Pp\Pp \to \PW\cPZ) = 30.7 \pm 9.3\stat \pm 7.1\syst \pm 4.1\thy \pm 1.0\lum\unit{pb}$ and $\sigma (\Pp\Pp \to \cPZ\cPZ) = 6.5 \pm 1.7\stat \pm 1.0\syst \pm 0.9\thy \pm 0.2\lum\unit{pb}$, are consistent with next-to-leading order quantum chromodynamics calculations.}

\hypersetup{%
pdfauthor={CMS Collaboration},%
pdftitle={Measurement of WZ and ZZ production in pp collisions at sqrt(s) = 8 TeV in final states with b-tagged jets},%
pdfsubject={CMS},%
pdfkeywords={CMS, physics, diboson}}

\maketitle 

\section{Introduction}
The study of WZ and ZZ (referred to collectively as VZ) diboson production in proton-proton collisions provides an important
test of the gauge sector of the standard model (SM). In \Pp\Pp\ collisions at $\sqrt{s} = 8\TeV$, the predicted
cross sections are \theoryXSWZ\ and
\theoryXSZZ\ at next-to-leading order (NLO) in quantum chromodynamics (QCD)~\cite{Campbell:2010ff}.
A significant deviation from these theoretical values would indicate contributions from physics beyond the SM.
Both processes constitute important backgrounds to the associated production of V and standard model Higgs (H) bosons,
especially in those channels involving $\PH\to\bbbar$ decays.
The production rate of two vector bosons in \Pp\Pp\ collisions at the Large Hadron Collider (LHC) has been measured by
the ATLAS and Compact Muon Solenoid (CMS) Collaborations
in all-leptonic WZ and ZZ decay modes~\cite{ATLASWZ, ATLASZZ, CMSWZ, cmsVV4l}.

We present a measurement of the VZ production cross sections
in the $\cV\cPZ\to \Vbb$ decay mode, where the V decays leptonically:
\ZToNN, \WToLN, and \ZToLL, with $\ell$ corresponding to either $\Pe$ or $\mu$.
Contributions from \WtoTN\ with leptonic $\tau$ decays are included in the \WToLN\ channels.
The analysis uses final states with no charged leptons (0-lepton), single lepton (1-lepton), or dilepton (2-lepton) events
with electron and muon channels analyzed separately.
The Z boson decays to \cPqb\ quarks are selected by requiring the presence of two b-tagged jets.
The results are based on data corresponding to an integrated luminosity of 18.9\fbinv collected
with the CMS detector at the LHC.
Two methods are used in the analysis, one involves a fit to the output of a multivariate discriminant,
and the other a fit to the two-jet mass (\mbb) distribution.
The cross sections are calculated simultaneously for WZ and ZZ production at transverse momenta
of the accompanying V of $\ptV > 100\GeV$, for Z boson masses
falling within the window $60<\MZ<120\GeV$. The latter requirement assures a
uniform treatment of interference with background processes.
Approximately 15\% of the WZ and 14\% of the ZZ total inclusive cross sections are
contained within their respective regions of acceptance for $\ptV > 100\GeV$,
as calculated using several event
generators discussed in the following section.
The \WtoLN\ channel is sensitive almost exclusively to WZ production, while
the \ZtoLL\ modes are restricted to the ZZ process.
The channel with no charged leptons is sensitive to both production modes, with ZZ and
WZ channels contributing 70\% and 30\%, respectively, to these events.
The 0-lepton WZ events contribute primarily when the lepton from \WToLN\ falls outside of acceptance.

\section{CMS detector, triggering,  object reconstruction and event simulation}

A description of the CMS detector can be found in Ref.~\cite{CMSDETECTOR}.
Particles produced in $\Pp\Pp$ collisions are detected in the pseudorapidity range $ \abs{\eta}< 5$,
where $\eta = -\ln[\tan(\theta/2)]$, and $\theta$ is the polar angle relative to the direction
of the counterclockwise circulating proton beam.
The CMS detector comprises a superconducting solenoid, providing a uniform axial magnetic
field of 3.8\unit{T} over a cylindrical region that is 12.5\unit{m} long and 6\unit{m} in diameter.
The magnetic volume contains a silicon pixel and strip tracking system  ($\abs{\eta} < 2.5$),
surrounded by a lead tungstate crystal electromagnetic  calorimeter (ECAL) and a brass/scintillator
hadronic calorimeter (HCAL) at $\abs{\eta} < 3.0$.
A steel/quartz-fiber Cherenkov calorimeter extends the coverage to $ \abs{\eta} = 5$.
The steel flux-return yoke outside the solenoid is instrumented with gas-ionization detectors used to
identify muons at $ \abs{\eta} < 2.4$.

The \WtoLN\ channels rely on several single-lepton
triggers with \pt\
thresholds between 17 and $30\GeV$ and restrictive lepton identification.
The 2-lepton channels use the same single-muon triggers for selecting the  \ZtoMM events and
2-electron triggers with \pt\ thresholds of 17 and 8\GeV for the electron of higher and lower \pt, respectively,
and with more restrictive isolation requirements for selecting the \ZtoEE events.

A combination of several triggers is used for the events without charged leptons:
all triggers require \MET to be
above a given threshold, such that the trigger efficiency ranges from 70\% to 99\% for $\MET=100\GeV$
 to $170\GeV$, respectively.

Electron reconstruction requires a match of
a cluster in the ECAL to a track reconstructed in the silicon
tracker~\cite{CMS-PAS-EGM-10-004, CMS-DP-2011-003, CMS-DP-2013-003}.
Electron identification relies on a multivariate technique that combines observables sensitive to the amount of bremsstrahlung emitted along the electron trajectory, the match in position and energy of the electron trajectory with the associated cluster, as well as the energy distribution in the cluster. Additional requirements are imposed to minimize background from electrons
produced through photons converting into $\Pep\Pem$ pair while traversing the tracker material. Electron candidates are
considered if observed in the pseudorapidity range $\abs{\eta} < 2.5$ but
excluding the transition regions at $1.44 < \abs{\eta}< 1.57$  between the ECAL barrel and endcaps.

Muons are reconstructed using two algorithms~\cite{Chatrchyan:2012xi}: one in which
tracks in the silicon tracker are matched to signals in the muon
chambers, and another in which a global fit is performed to the track that is seeded by
signals detected in the outer muon system.  The muon
candidates are required to be reconstructed by
both algorithms.  Additional identification criteria  are
imposed on muon candidates to reduce the fraction
of tracks misidentified as muons. These include the number of hits reconstructed in the tracker and
in the muon system,
the quality of the global fit to a muon trajectory, and its consistency with originating from the primary
vertex. Muon candidates are finally required to fall in the $\abs{\eta} < 2.4$ range.

Jets are reconstructed from particle-flow~\cite{CMS-PAS-PFT-09-001,CMS-PAS-PFT-10-002} objects using the anti-\kt jet clustering algorithm~\cite{antikt}, with a distance parameter of 0.5,
as implemented in the \textsc{fastjet}
package~\cite{Cacciari:fastjet1,Cacciari:fastjet2}.  Each jet is required to
lie within $\abs{\eta} < 2.5$ and have $\pt > 20\GeV$.
Jet energy corrections are applied as a function of $\eta$ and $\pt$ of the jet~\cite{cmsJEC}.
The imbalance in transverse momentum (often referred to as ``missing transverse energy vector'') is calculated as the negative
of the vectorial sum of the $\ptvec$ of all particle-flow objects identified in the
event, and the magnitude of this vector is referred to as \MET.
The procedures of Ref.~\cite{VHbb} are applied on an event-by-event basis
to mitigate the effects of multiple interactions per beam crossing (pileup).

The CMS combined secondary-vertex (CSV) b-tagging
algorithm~\cite{Chatrchyan:2012jua} is used to identify jets that are likely to
originate from the hadronization of b quarks.  This algorithm combines
the information about track impact parameters and secondary
vertices in a discriminant that distinguishes
 \cPqb\ jets from jets originating from light quarks, gluons, or
\cPqc\ quarks. The output of the CSV algorithm is a continuous discriminator with a value in the range 0 to 1,
where typical thresholds for \cPqb\ jet selection range from loose ($\approx$0.2) to tight ($\approx$0.9).
Depending on the chosen CSV threshold, the efficiencies for tagging jets originating from \cPqb\ quarks range from 50\% (tight)
to 75\% (loose), while the misidentification rates for \cPqc\ quarks range from 5\% (tight) to 25\% (loose)
and for light quarks or gluons range from 0.2\% (tight) to 3\% (loose).

The b-jet energy resolution
is improved by applying multivariate regression techniques similar to those used in
the CDF experiment~\cite{1107.3026}. An additional correction, beyond the standard CMS jet energy corrections,
is derived from simulated events to recalibrate each b-tagged jet with the generated \cPqb\ quark energy.
This involves a specialized boosted decision tree (\BDT )~\cite{Roe:2004na,Hocker:2007ht} trained on
simulated signal events, with
inputs that include information on jet structure,
such as information about individual tracks, jet constituents, information on semileptonic b-hadron decays, and
the presence of any low-\pt leptons. The \BDT\ correction, identical to that used in Ref.~\cite{VHbb},
improves the resolution on the mass of the \bbbar\ system by ${\approx}$15\%,
resulting in an increase in the sensitivity of the analysis of
10--20\%, depending on the specific channel.
The \ZtoBB\ invariant mass resolution
after this correction is ${\approx}$10\%.

Simulated samples of events are produced using
several event generators, and the response of the CMS detector is modeled
using the \GEANTfour program~\cite{GEANT4}.
The {\MADGRAPH 5.1}~\cite{Alwall:2011uj} generator is used to generate
the diboson signals, as well as the background from
W+jets, Z+jets, and \ttbar\ events.
The single-top-quark samples are generated with \POWHEG~\cite{Nason:2004rx,POWHEG,Alioli:2010xd,Alioli:2009je}, and generic
multijet samples using {\PYTHIA 6.4}~\cite{PYTHIA}.
VH event samples with a SM H boson mass of $\mH = 125\GeV$ are also
produced using the \POWHEG~\cite{Nason:2009ai} event generator
interfaced to {\HERWIG++}~\cite{Gieseke:2006ga} for parton showering and hadronization.
The NLO {MSTW2008} set~\cite{Martin:2009iq} of parton distribution
functions (PDF) is used to produce the NLO \POWHEG samples, while the leading-order (LO)
CTEQ6L1 set~\cite{CTEQ6} is used for the events that correspond to
LO calculations. The Z2Star tune~\cite{Chatrchyan:2011id}
is used to parametrize the underlying event.
Corrections to account for differences in efficiencies between data and simulation are
measured using data using a tag and probe technique~\cite{CMSTNP},
and applied as individual weights to each of the simulated events.

\section{Event selection}
We use the analysis techniques developed in the CMS VH studies of Ref.~\cite{VHbb}.
Event selection is based on the
reconstruction of a vector boson that decays leptonically in association with the Z boson that decays
into two b-tagged jets.
Dominant backgrounds to VZ production include V+\cPqb\ jets, V+light flavor
(LF = \cPqu\cPqd\cPqs\cPqc\ quark or gluon) jets,
\ttbar, single-top-quark, generic multijet, and H boson production.
In general, b-tagging reduces the contributions from LF events, and counting additional jet activity
is used to reduce background from \ttbar and
single-top-quark events. Finally, the value of \mbb\
provides a way to distinguish VZ from V+\cPqb\ and SM VH production,
as discussed below.

The reconstruction of a \ZtoBB\ decay proceeds by selecting two
central jets from the primary vertex with $\abs{\eta}<2.5$, each with a \pt above some chosen
threshold, and defining the \bbbar\ candidate as the jet pair with largest vectorial sum
of transverse momenta (\ptjj). This combination is very efficient for $\ptV>100\GeV$
without biasing the differential distribution of the background, and also defines the two-jet mass \mbb,
which is required to be $<250\GeV$. The two selected jets are also required to
be tagged as \cPqb\ jets, with a value of the CSV discriminator
that depends on the specific nature of the event.

Candidate \WToLN\ decays in WZ events are identified through
the presence of a single isolated lepton and significant \MET.
Electrons and muons are required to have $\pt>30\GeV$ and $\pt>20\GeV$, respectively.
To reduce contamination from generic multijet processes, the \MET is required to be $>45\GeV$. In addition,
the azimuthal angle ($\phi$) between the \MET\ vector and the lepton
is required to be $<\pi/2$. At least two jets with $\pt > 30\GeV$
and a moderate CSV discriminator value are required to define the \ZtoBB\ candidate.

Candidate \ZToLL\ decays in ZZ events are reconstructed by combining
isolated, oppositely charged pairs of electrons or muons, with a dilepton invariant
mass of $75<m_{\ell\ell}<105\GeV$.  The
\pt\ of each lepton is required to be $>20\GeV$.
The two jets of the \ZtoBB\ candidate must pass a loose CSV discriminator value,
which is optimized in simulated events for increasing the sensitivity of the analysis.

The identification of \ZToNN\ decays in ZZ events
requires $\MET >100\GeV$ in the event, and
at least one of the \cPqb\ jets with $\pt > 60\GeV$ and the other with  $\pt > 30\GeV$
to form a \ZtoBB\ candidate.
Moderate CSV requirements are applied on both jets.
Two additional event requirements are imposed to
reduce the multijet background in which \MET originates from
mismeasured jet energies.
First, a $\dphiMJ>0.5$ radians requirement is applied on the azimuthal angle
between the direction of \MET and the \pt of the jet
closest in $\phi$ that satisfies $\abs{\eta}<2.5$ and $\pt>25\GeV$.
 The second requirement is that the
 azimuthal angle between the direction of ${\MET}^\text{(trks)}$, as
calculated from only the charged tracks that satisfy $\pt>0.5\GeV$ and $\abs{
\eta}<2.5$, and the direction of the full \MET has $\dphiMtrk <0.5$ radians.
Finally, to reduce background from \ttbar\ events in the \WtoLN\
and \ZtoNN\ channels, events that contain any additional isolated leptons with $\pt>20\GeV$ are
rejected.

\subsection{Multivariate analysis}\label{sec:hbb_BDT}

The signal region is defined by events that satisfy the V and Z boson reconstruction criteria
described above. To optimize the significance of the signal as well as the \bbbar\ mass resolution,
events are classified into different regions of the V boson transverse momentum.
In particular, we define three regions for the \WtoLN\ channels:
(i)~$100<\ptV<130\GeV$, (ii) ~$130<\ptV<180\GeV$, and (iii)~
$\ptV>180\GeV$. A single inclusive region of $\ptV>100\GeV$ is defined for the \ZtoLL\ channels.
Three regions for the channel without charged leptons are defined by (i)~$100<\ptV<130\GeV$, (ii)~$130<\ptV<170\GeV$, and
(iii)~$\ptV>170\GeV$. For regions (i) and (ii), the requirement on \dphiMJ\ is
tightened to $\dphiMJ>0.7$ radians. To reduce background in the region of smallest \ptV,
 the \MET significance (defined as the ratio of \MET to
the square root of the total transverse energy deposited in the
calorimeter) is required to be ${>}3 \sqrt{\smash[b]{\GeVns}}$.

To better discriminate between signals and background, the final stage of the analysis introduces
a BDT discriminant trained on simulated samples for signal and all background
processes. The set of input variables is identical to the one used in Ref.~\cite{VHbb}, and includes the mass of the \bbbar\ system,
the number of additional jets beyond the \cPqb\ and \cPaqb\ candidates (\Naj), the value of CSV for the \bbbar\ jets with \CSVmin\
specifying the smaller value and \CSVmax\ the larger one,
and the distance in $\eta$-$\phi$ between the \cPqb\ and \cPaqb\ jet axes, $\dRJJ = \sqrt{\smash[b]{\left(\Delta \phi \right)^2+\left(\Delta \eta \right)^2}}$.

\begin{figure}[tbp]
\centering
{\includegraphics[width=0.49\textwidth]{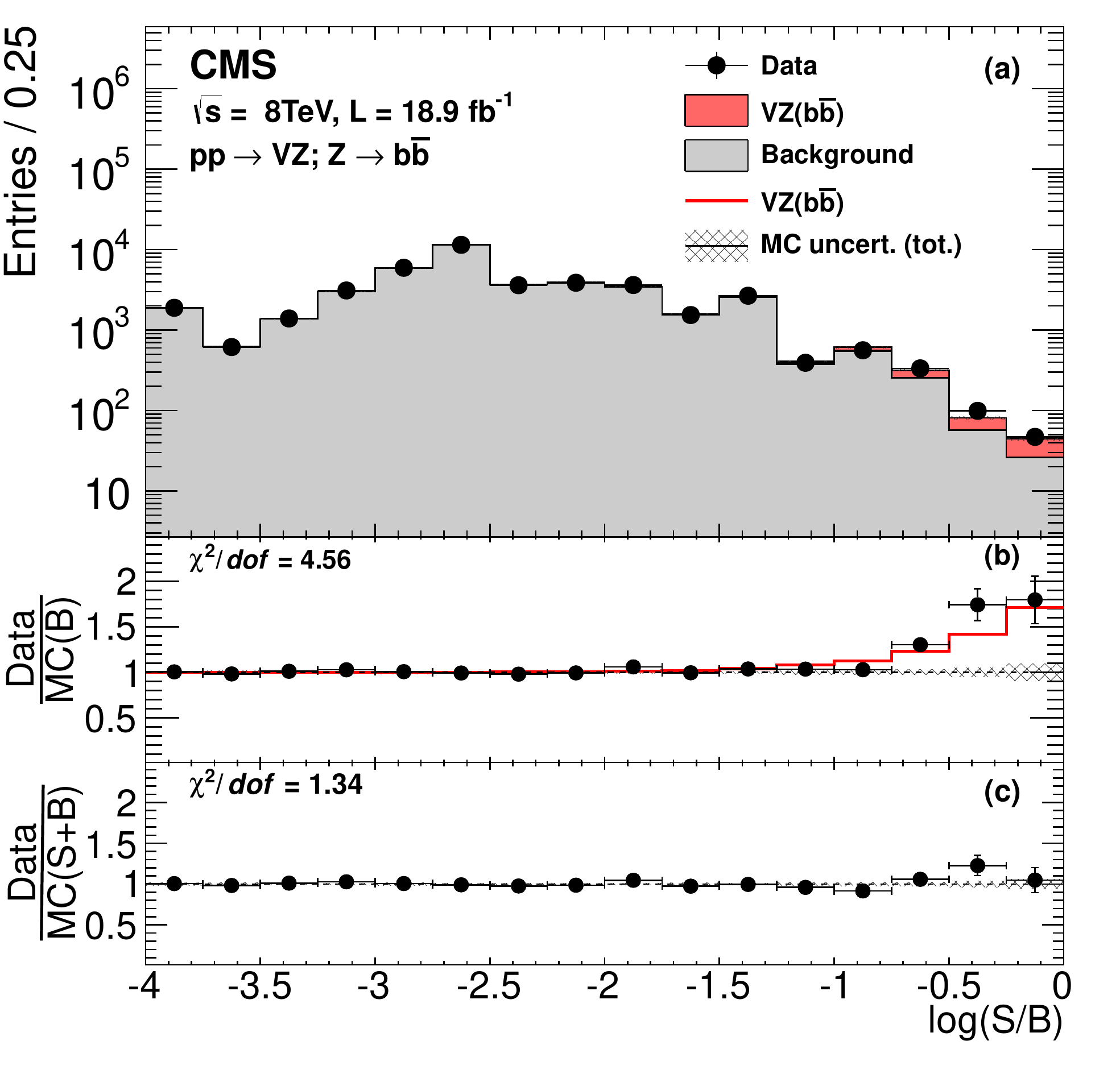}}
\caption{(a) Combined distribution for all channels in the value of the logarithm of the ratio of
    signal to background (S/B) discriminants in data and in Monte Carlo (MC) simulations,
    based on the outputs of the S and B BDT discriminants for each event.
    The two bottom panels display (b) the ratio of the data and of the SM expectation relative
    to the background-only hypothesis, and (c) data relative to the expected sum of background and VZ signal.
    The error bars and the cross-hatched regions reflect total uncertainties at 68\% confidence level.}
   \label{fig:BDT_S_over_B_all}
\end{figure}

Figure~\ref{fig:BDT_S_over_B_all}(a) displays the combined differential distribution for events
from all channels as a function of the logarithm
of the signal-to-background (S/B) ratio
of the values of the output of the corresponding S and B contributions to the \BDT\
discriminants of each event. Panel~(b) gives the ratio
of the data (black points) to the SM expectation (histogram) relative
to the background-only hypothesis, while panel~(c) gives the ratio
to the expectation from the SM, including the VZ contribution.
The excess
observed in bins with largest
S/B is clearly consistent with what is expected for
VZ production in the SM.

\subsection{Two-jet mass analysis}\label{sec:hbb_Mjj}

As a cross-check of the multivariate analysis, we perform a simpler analysis based
on the \mbb\ distribution of the reconstructed \bbbar\ jets of the hypothesized Z boson.
The signal region is defined by events that satisfy the V and Z boson reconstruction criteria
used in the multivariate analysis. Events are again classified according to \ptV, and, in addition,
more restrictive selections are introduced than in the multivariate
analysis, because the single variable \mbb\ is not a sufficiently sensitive discriminant.

In the \ZtoNN\ and \WtoLN\ channels,
the b-tagging requirements are tightened, respectively, to a tight
$\CSVmax$ and a medium $\CSVmin$. A veto
is also imposed on any additional jets,
and $\dphiVZ$ is required to be ${>}2.95$ radians.
The regions of $100<\ptV<130\GeV$,  $130<\ptV<180\GeV$, and
$\ptV>180\GeV$ are used to analyze the 1-muon channel, and the
regions for the 1-electron channel are defined as
$100<\ptV<150\GeV$ and $\ptV>150\GeV$.
The selected regions for the \ZtoNN\ channel
are identical in \ptV\ to the requirements used in the multivariate analysis,
but we define ranges of $\ptjj>110\GeV$,  $\ptjj>140\GeV$, and
$\ptjj>190\GeV$, and impose an additional threshold for the jet of highest \pt of
${>}80\GeV$ for the region of $\ptjj>140\GeV$.
For the \ZtoLL\ channels, the \ptV\
ranges are defined by $100<\ptV<150\GeV$
and $\ptV>150\GeV$, and, in addition, we require medium
$\CSVmax$ and moderate $\CSVmin$ thresholds,
and $\MET < 60\GeV$.

\begin{figure}[tbp]
 \centering
   \includegraphics[width=0.49\textwidth]{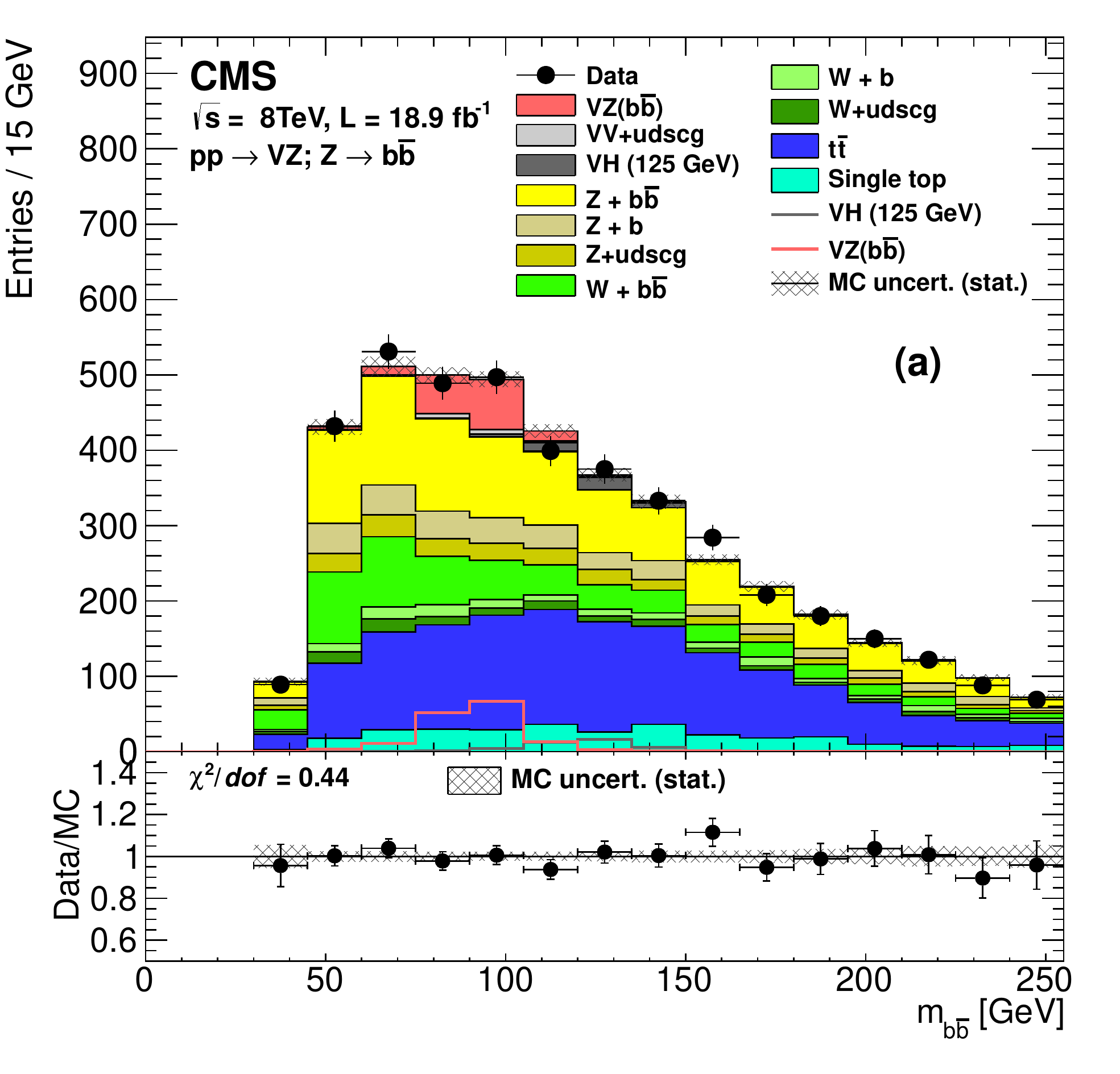}
   \includegraphics[width=0.49\textwidth]{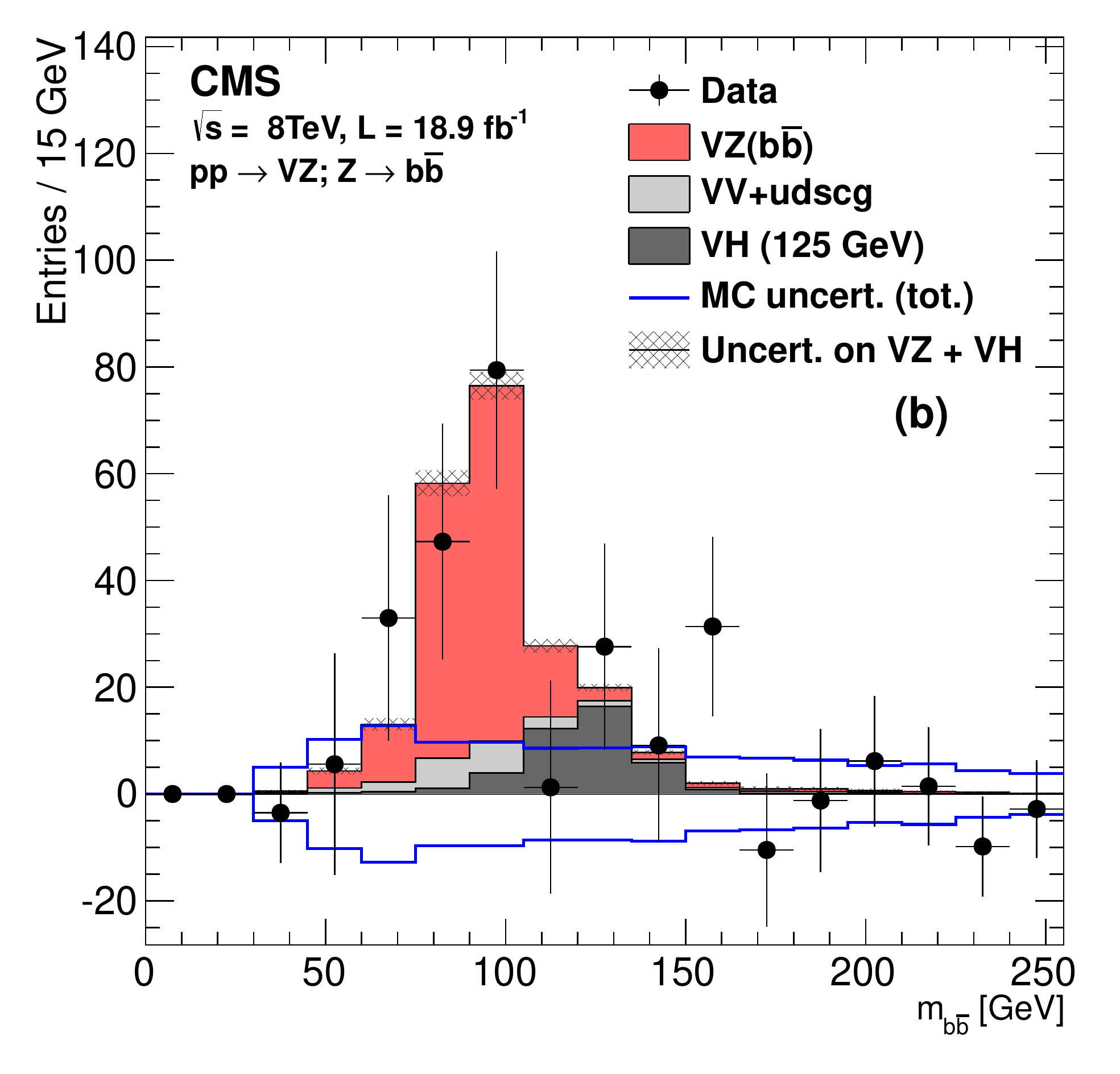}
   \caption{(a) The combined \bbbar\ invariant mass distribution for all
       channels, compared to MC simulation of SM contributions. (b) Same distribution as in (a), but with all backgrounds to
VZ production, except for the VH contribution, subtracted. The contributions from
backgrounds and signal are summed cumulatively. The expectations for the sum of VZ signal and background from
VH production are also shown superimposed. The error bars and cross-hatched regions reflect statistical uncertainties at 68\% confidence level.}
   \label{fig:MJJ-combined}
\end{figure}

Figure~\ref{fig:MJJ-combined}~(a)
combines events from all channels into a single \mbb\ distribution,
which is compared to expectations from the SM.
Figure~\ref{fig:MJJ-combined}~(b) shows the same distribution, but
after subtracting all SM contributions except for the VZ signals and VH backgrounds.
The VZ signal is clearly
visible, with a yield compatible to that expected in the SM.

\section{Background calibration regions and systematic uncertainties}

Calibration regions in data are used to
validate the simulated distributions used to build the \BDT\
discriminants, as well as to correct normalizations of the major background contributions from
W and Z bosons produced in association with jets (LF or \cPqb\ quarks) and \ttbar\ production.
These calibration regions are identical to those of Ref.~\cite{VHbb}, and typically involve inversion
of b-tag selection criteria and two-jet mass sidebands around the signal region.
A set of simultaneous fits is then performed to distributions of discriminating variables
in the calibration regions, separately for
each channel, to obtain consistent scale factors that are used to adjust the yields from simulated events.  These scale factors
account not only for discrepancies between predicted cross sections and data, but also for any residual
differences in the selection of physical objects.
Separate scale factors are consequently applied for each of the background
processes in the different channels.
For the backgrounds from V+jets, the
calibration regions are enriched in either
\cPqb\ or LF jets. Uncertainties in the scale factors include
statistical components arising from the fits to the discriminant
(affected by the finite size of the data and MC samples),
and systematic uncertainties originating from \cPqb\ tagging, jet energy scale, and jet energy resolution.
The numerical values of the scale factors are close to unity and their uncertainties (3--50\%)
are identical to those of Ref.~\cite{VHbb}.

The systematic uncertainties considered in the measurement of the cross section
using the multivariate analysis are summarized in Table~\ref{tab:syst}.
The two columns give the uncertainty in the ``signal strength'' $\mu$ for the WZ and ZZ processes,
which corresponds to the ratio of the observed yield relative to the
yield expected from the SM. Each systematic uncertainty is represented by a nuisance parameter
and profiled in the combined fit. To evaluate the impact of individual uncertainties
a fit to a simulated pseudo-dataset is performed removing individual nuisance parameters.

Theoretical uncertainties in the acceptances are evaluated using the
{\MCFM}~\cite{Campbell:2010ff} generator
by changing the QCD factorization and renormalization scales up and down by a factor of two relative to the default scales
of $\mu_R = \mu_F = m_Z$.
The impact of uncertainties in PDF and $\alpha_s$ on the cross section and acceptance
of the VZ signal are evaluated following the PDF4LHC prescription~\cite{Botje:2011sn,Alekhin:2011sk},
using CT10~\cite{Lai:2010vv}, MSTW08~\cite{Martin:2009iq}, and NNPDF2.0~\cite{Ball:2010de} sets of PDF,
and the combined uncertainty is found to be 5\% for both WZ and
ZZ production. Because of the large \ptV\ values required in this analysis,
the results are sensitive to electroweak (EW) and  NNLO QCD corrections, both of which can be significant.
Since the exact corrections for the VZ process are not available,
we use the NLO EW~\cite{HAWK1,HAWK2,HAWK3} and
next-to-next-to-leading-order (NNLO) QCD~\cite{Ferrera:2011bk}
corrections to VH production, and apply these to the VZ channel,
because they are expected to be similar for the two processes.
Based on the size of the correction, an additional 10\% uncertainty
is assigned to the inclusive cross section to account for the extrapolation
to the $\ptV<100\GeV$ region.

The uncertainty in CMS luminosity is estimated to be 2.6\%~\cite{CMS-PAS-LUM-13-001}.
Muon and electron triggering,
reconstruction, and identification efficiencies are
determined in data from samples of
\ZToLL\ decays. The uncertainty in the lepton yields due to trigger inefficiency is
2\% per lepton, as is the uncertainty in lepton
identification efficiency. The parameters describing the turn-on in the
trigger efficiency in the \ZtoNN\ channel are varied
within their statistical uncertainties for different
assumptions on the methods used to derive the efficiency. The estimated
uncertainty is 3\%.

The jet energy scale is also varied
within its uncertainty as a function of jet \pt\ and
$\eta$, and the efficiency of the selections is then
recomputed to assess the dependence on these variables. The effect of this uncertainty
on the jet energy resolution is evaluated by
smearing the jet energies according to their measured
uncertainties, a process that affects both the normalization and distribution of events. An uncertainty of 3\% is assigned to
the yields of all processes in the \ZtoNN\ and \WtoLN\ channels due to
uncertainties related to \MET, such as its scale and resolution.

Scaling factors to normalize b-tagging in simulation to that in data (measured in \cPqb\
enhanced samples of jets that contain muons)
are applied  consistently to jets in simulated signal and background
events. The measured uncertainties in b-tagging scale factors are
3\% per b-quark jet, 6\% per c-quark jet, and 15\% per
mistagged jet (originating from a gluon or from a light quark)~\cite{Chatrchyan:2012jua}.
These translate into uncertainties in yields of 3--15\%, depending on channel and specific process. The
\BDT\ output is also affected by the distributions of the CSV output,
and an uncertainty is therefore assigned according to ${\pm}1$ standard deviation (SD)
variation in yield and shape of the CSV distributions.

Finally, the sizes of the simulated samples, as well as uncertainties in
generator-level modeling of V+jets and \ttbar\ backgrounds,
are taken into account to determine the total uncertainty in the signal strength $\mu$.

\begin{table*}[tbp]
\topcaption{Sources of systematic uncertainty,
    including whether they affect the distribution (dist) or normalization (norm) of the
    \BDT\ output,
    and their relative contributions to the expected uncertainty
    in the signal strengths $\mu_{\PW\cPZ}$ and $\mu_{\cPZ\cPZ}$ after fitting the model.
    }
    \label{tab:syst}
    \newdimen\twid\setbox0=\hbox{Individual contributions to uncertainty}\dimen0=\wd0\divide\dimen0 by 2\twid=\dimen0
    \centering
    \begin{tabular}{lc|D{.}{.}{-1}D{.}{.}{-1}} \hline
        &           & \multicolumn{2}{c}{\unhbox0} \\
        Source of uncertainty &    Type   &  \multicolumn{1}{c}{$\mu_{\PW\cPZ}$ (\%)}  &  \multicolumn{1}{c}{$\mu_{\cPZ\cPZ}$ (\%)}  \\  \hline
Luminosity                                    & norm   & 3.3     & 3.2  \\
Lepton efficiency and trigger                 & norm   &1.9               & 0.6    \\
\ZtoNN\ triggers                              & dist           &\multicolumn{1}{c}{---}               & 1.6     \\
Jet energy scale                              & dist           &7.2    & 6.4     \\
Jet energy resolution                         & dist           &6.1          & 5.9      \\
\MET                                          & dist           &3.3               & 1.8    \\
b tagging                                     & dist           &7.7         & 5.7  \\
VZ cross section (theory)                     & norm   &13.4               & 13.4     \\
Monte Carlo statistics                        & dist           &5.5           & 3.6      \\
Backgrounds (from data)                       & norm   &12.5  & 11.5     \\
Single-top and VH (from simulation)           & norm   &1.9              & \multicolumn{1}{c}{---}     \\
MC modeling of V+jets and \ttbar              & dist           &4.7              & 4.8  \\[-2ex]
&&\multicolumn{1}{c}{\rule{\twid}{0pt}}&\\
\hline
\end{tabular}
\end{table*}

\section{Results}

The total cross sections are determined from a simultaneous fit to all final states,
constrained by the number of events observed in each category.
The likelihood is written as a combination of
individual channel likelihoods for
the signal and background hypotheses.
We extract the best-fit values of the signal strength assuming the SM expectation for the ratio of
$\sigma{(\PW\cPZ)}/\sigma{(\cPZ\cPZ)}$ at NLO.
Using the baseline multivariate analysis, the VZ process is observed with a statistical significance of
6.3 SD (5.9 SD expected). The measurement corresponds
to a signal strength relative to the SM of \muBDT.
The cross-check analysis based on \mbb\ yields a significance of 4.1 SD (4.6 SD expected), which corresponds to \muMbb.
In the following, the interpretation refers to the more sensitive multivariate analysis.

\begin{figure}[tbp]
\centering
{\includegraphics[width=0.45\textwidth]{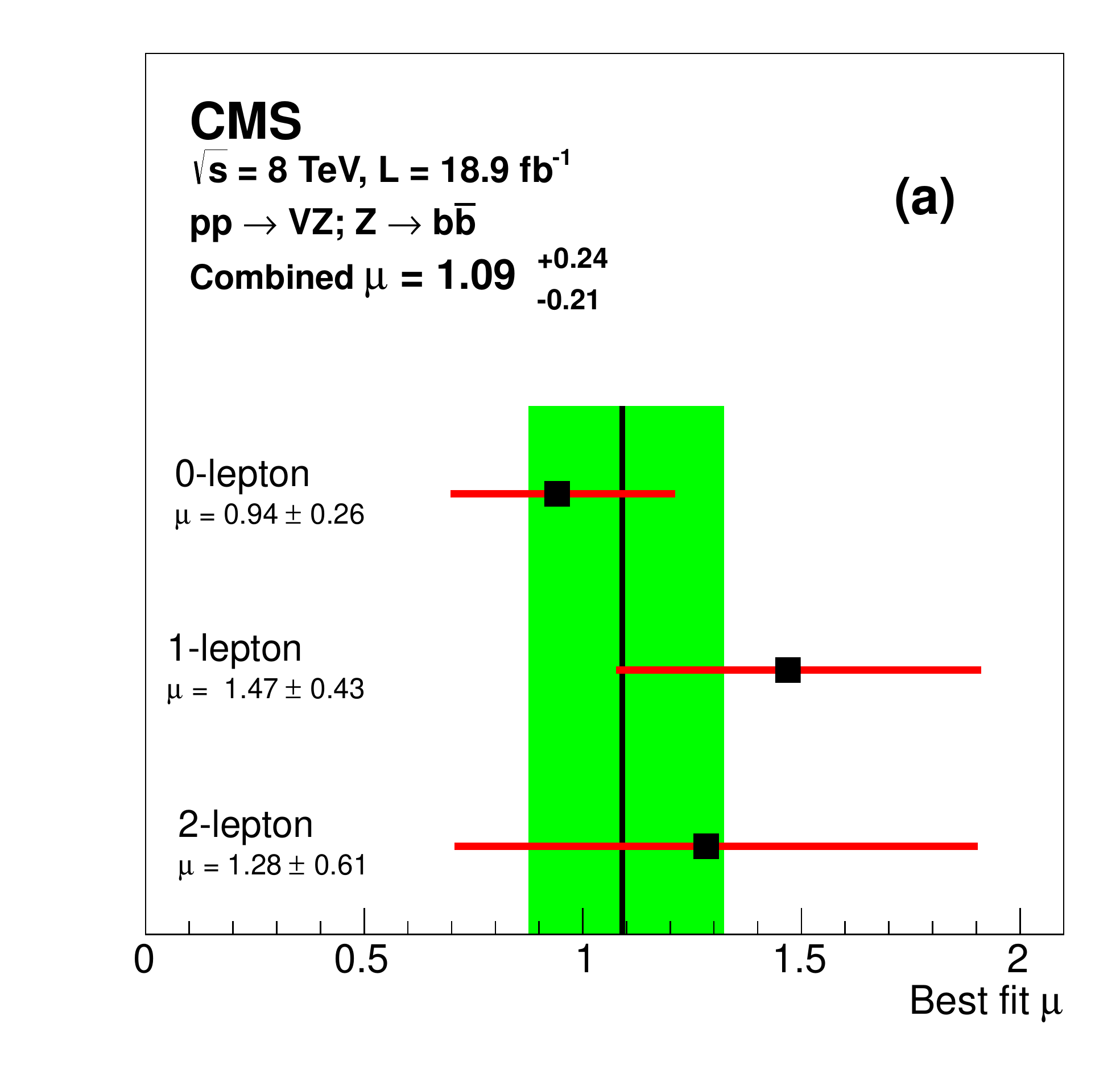}}
{\includegraphics[width=0.45\textwidth]{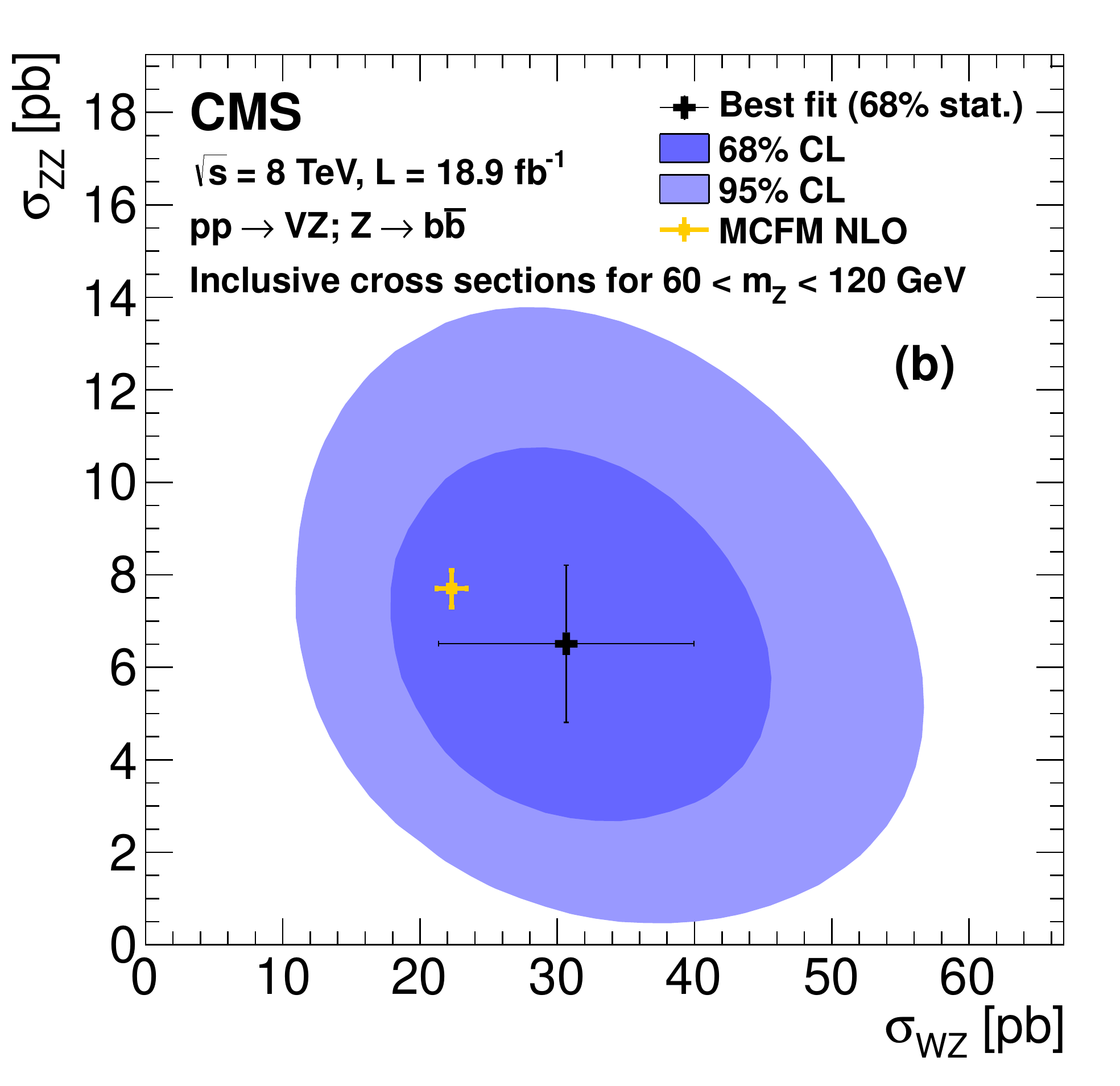}}
\caption{(a) Best-fit values of the ratios of the VZ production cross sections,
relative to SM predictions for individual channels, and for all channels combined (hatched band). (b)
Contours of 68\% and 95\% confidence level for WZ and ZZ production cross sections. The large cross indicates the best-fit
value including its 68\% statistical uncertainty, and the light small cross shows the result for the MCFM NLO calculation.}
   \label{fig:SignalStrength}
\end{figure}

The cross sections extracted from the individual channels are consistent with each other and with the SM predictions,
as can be seen in Fig.~\ref{fig:SignalStrength}~(a). To extract the WZ and ZZ
cross sections, a simultaneous fit is performed floating independently the WZ and ZZ contributions,
with results displayed in Fig.~\ref{fig:SignalStrength}~(b).
The most likely values are \muWZ\
and \muZZ.

The values for the signal strength are extrapolated to the mass window $60<\MZ<120\GeV$
for both the \bbbar\ and lepton pair invariant masses.
The resulting cross section for inclusive WZ production is
\XSWZ, compared to the theoretical value of
\theoryXSWZ,
 calculated with \MCFM using the MSTW2008 PDF.
The ZZ cross section is
\XSZZ, for the same Z-mass window, which
can be compared to the theoretical value of
\theoryXSZZ, also
 calculated with \MCFM
using the MSTW2008 PDF.
The uncertainties in both theoretical values
include uncertainties
in the PDF and $\alpha_s$, and those
originating from the uncertainty in renormalization and factorization scales.
The ZZ cross section is in agreement with CMS measurements using all-leptonic V decays
of Ref.~\cite{cmsVV4l}, which is more precise
than this analysis.

{\tolerance=800
The cross sections for $\ptV > 100\GeV$ and for Z bosons produced
in the mass region $60<\MZ<120\GeV$
are determined to be \XSWZfid\ and \XSZZfid. The acceptance for this $\pt$ region has smaller theoretical uncertainty,
estimated as 1\% using MC signal simulation; the measurements are found in agreement with
the NLO \MCFM calculations yielding \theoryXSWZfid\ and \theoryXSZZfid.
\par
}
\section{Summary}

We presented measurements of the inclusive $\Pp\Pp \to\cV\cPZ$ (where V denotes W or Z) cross sections
in data recorded by the CMS experiment at the LHC at $\sqrt{s} =8\TeV$,
corresponding to an integrated luminosity of
18.9\fbinv. The measurements are based on $\cV\cPZ\to \Vbb$ final states.
The decay modes \ZToNN, \WToLN, and \ZToLL\ ($\ell = \Pe, \Pgm$) are used
to identify the accompanying V. We observe $\cV\cPZ\to \Vbb$ production with a combined significance of 6.3 standard deviations.
The total cross sections, defined for $60<\MZ<120\GeV$, are found to be
\XSWZ\ and \XSZZ. These values are consistent with the predictions \theoryXSWZ\ and \theoryXSZZ\
of the standard model at next-to-leading order.

\section*{Acknowledgments}

{\tolerance=1200
We congratulate our colleagues in the CERN accelerator departments for the excellent performance of the LHC and thank the technical and administrative staffs at CERN and at other CMS institutes for their contributions to the success of the CMS effort. In addition, we gratefully acknowledge the computing centres and personnel of the Worldwide LHC Computing Grid for delivering so effectively the computing infrastructure essential to our analyses. Finally, we acknowledge the enduring support for the construction and operation of the LHC and the CMS detector provided by the following funding agencies: BMWF and FWF (Austria); FNRS and FWO (Belgium); CNPq, CAPES, FAPERJ, and FAPESP (Brazil); MES (Bulgaria); CERN; CAS, MoST, and NSFC (China); COLCIENCIAS (Colombia); MSES and CSF (Croatia); RPF (Cyprus); MoER, SF0690030s09 and ERDF (Estonia); Academy of Finland, MEC, and HIP (Finland); CEA and CNRS/IN2P3 (France); BMBF, DFG, and HGF (Germany); GSRT (Greece); OTKA and NIH (Hungary); DAE and DST (India); IPM (Iran); SFI (Ireland); INFN (Italy); NRF and WCU (Republic of Korea); LAS (Lithuania); MOE and UM (Malaysia); CINVESTAV, CONACYT, SEP, and UASLP-FAI (Mexico); MBIE (New Zealand); PAEC (Pakistan); MSHE and NSC (Poland); FCT (Portugal); JINR (Dubna); MON, RosAtom, RAS and RFBR (Russia); MESTD (Serbia); SEIDI and CPAN (Spain); Swiss Funding Agencies (Switzerland); NSC (Taipei); ThEPCenter, IPST, STAR and NSTDA (Thailand); TUBITAK and TAEK (Turkey); NASU and SFFR (Ukraine); STFC (United Kingdom); DOE and NSF (USA).

Individuals have received support from the Marie-Curie programme and the European Research Council and EPLANET (European Union); the Leventis Foundation; the A. P. Sloan Foundation; the Alexander von Humboldt Foundation; the Belgian Federal Science Policy Office; the Fonds pour la Formation \`a la Recherche dans l'Industrie et dans l'Agriculture (FRIA-Belgium); the Agentschap voor Innovatie door Wetenschap en Technologie (IWT-Belgium); the Ministry of Education, Youth and Sports (MEYS) of Czech Republic; the Council of Science and Industrial Research, India; the Compagnia di San Paolo (Torino); the HOMING PLUS programme of Foundation for Polish Science, cofinanced by EU, Regional Development Fund; and the Thalis and Aristeia programmes cofinanced by EU-ESF and the Greek NSRF.
\par}

\bibliography{auto_generated}   

\cleardoublepage \appendix\section{The CMS Collaboration \label{app:collab}}\begin{sloppypar}\hyphenpenalty=5000\widowpenalty=500\clubpenalty=5000\textbf{Yerevan Physics Institute,  Yerevan,  Armenia}\\*[0pt]
S.~Chatrchyan, V.~Khachatryan, A.M.~Sirunyan, A.~Tumasyan
\vskip\cmsinstskip
\textbf{Institut f\"{u}r Hochenergiephysik der OeAW,  Wien,  Austria}\\*[0pt]
W.~Adam, T.~Bergauer, M.~Dragicevic, J.~Er\"{o}, C.~Fabjan\cmsAuthorMark{1}, M.~Friedl, R.~Fr\"{u}hwirth\cmsAuthorMark{1}, V.M.~Ghete, C.~Hartl, N.~H\"{o}rmann, J.~Hrubec, M.~Jeitler\cmsAuthorMark{1}, W.~Kiesenhofer, V.~Kn\"{u}nz, M.~Krammer\cmsAuthorMark{1}, I.~Kr\"{a}tschmer, D.~Liko, I.~Mikulec, D.~Rabady\cmsAuthorMark{2}, B.~Rahbaran, H.~Rohringer, R.~Sch\"{o}fbeck, J.~Strauss, A.~Taurok, W.~Treberer-Treberspurg, W.~Waltenberger, C.-E.~Wulz\cmsAuthorMark{1}
% ...
\vskip\cmsinstskip
% ...
1:~~Also at Vienna University of Technology, Vienna, Austria\\
2:~~Also at CERN, European Organization for Nuclear Research, Geneva, Switzerland\\
% ...

\end{sloppypar}
\end{document}